\documentstyle[aas2pp4]{article}

\def\ecs{ergs cm$^{-2}$ s$^{-1} \ $}

\def\lae{\mathrel{<\kern-1.0em\lower0.9ex\hbox{$\sim$}}}
\def\gae{\mathrel{>\kern-1.0em\lower0.9ex\hbox{$\sim$}}}

\lefthead{Cagnoni et al.}
\righthead{A Hard Medium Survey with the ASCA GIS}

\begin{document}

\title{A Hard Medium Survey with the ASCA GIS:\\
 the (2-10 keV) Number Counts Relationship}
\author{I. Cagnoni\altaffilmark{1,2,3} , R. Della 
Ceca\altaffilmark{4,5} and T. Maccacaro\altaffilmark{4,6}}
\affil{$^{1}$Universit\`a degli Studi di Milano, Via Celoria 16, 
20133, Milano, Italy}
\affil{$^{2}$ Present address: Harvard-Smithsonian Center for Astrophysics, 
60 Garden Street, Cambridge, Massachusetts 02138, USA}
\affil{$^{3}$E-mail: ilaria@alessandro.harvard.edu}
\affil{$^{4}$Osservatorio Astronomico di Brera, Via Brera 28, 
20121 Milano, Italy}
\affil{$^{5}$E-mail: rdc@brera.mi.astro.it}
\affil{$^{6}$E-mail: tommaso@brera.mi.astro.it}

\begin{abstract}

In this paper we report the first results on a medium survey 
program conducted in the 2-10 keV energy band using data from the GIS2
instrument onboard the ASCA satellite.

We have selected from the ASCA public archive (as of February 14, 1996)
87 images which are suitable for this project. 
Sixty serendipitous X-ray sources, with a signal-to-noise ratio 
greater than 3.5, were found.
The 2-10 keV flux of the detected sources ranges from $\sim 1.1 
\times 10^{-13}$ \ecs to $\sim 4.1 \times 10^{-12}$  \ecs.

Using this sample we have extended the description of the 2-10 
keV LogN($>$S)--LogS to a flux limit of $\sim 6.3\times 10^{-14}$ \ecs
(the faintest detectable flux), i.e.  about 2.7 
orders of magnitude fainter than the Piccinotti et al. (1982) determination.  
The derived number-flux relationship is well described by a power 
law model, $N(>S) = K \times S^{-\alpha}$, with best fit values  
$\alpha = 1.67\pm 0.18$ and $K = 2.85\times 10^{-21}$ deg$^{-2}$.

At the flux limit of the survey about 
27 \% of the Cosmic X-ray Background in the 2-10 keV energy band is 
resolved in discrete sources.
A flattening of the number-flux relationship, within a factor of 10 from 
the flux limit of the present survey, is expected in order to avoid saturation.
  
The implications of these results on the models for the origin 
of the hard X-ray background are briefly discussed.

\end{abstract}


\section{Introduction}

The origin of the Cosmic X-ray Background (CXB), discovered 
almost 35 years ago (Giacconi et al., 1962), represents one of the 
long-standing problem of modern cosmology (see the reviews by
Fabian and Barcons, 1992;  Setti and Comastri, 1996).

The very small deviation from a blackbody shape of the cosmic 
microwave background spectrum  seems to have put to 
rest the suggestion that a significant fraction of the CXB is due to 
truly diffuse emission from a hot intergalactic medium (Mather et al., 1994).  
Therefore only the alternative interpretation, the discrete sources 
origin, is left.

X-ray surveys provide a powerful tool to study the nature
and properties of the  classes of X-ray emitters which produce the CXB.  
In the soft X-ray energy band ($E < 2$ keV), where grazing incidence
focusing optics have been used, deep surveys conducted with the 
ROSAT observatory (Hasinger et al., 1993; 
Branduardi-Raymont et al., 1994; Vikhilin et
al., 1995) have revealed a surface density of X-ray sources of 
$\sim 415$ deg$^{-2}$ at a flux limit  of $2.5\times 
10^{-15}$ \ecs (0.5 -- 2.0 keV), contributing $\sim$ 60-65 \% of the CXB in the same energy band.
The differential log(N)--log(S) can be described 
(see e.g. Hasinger et al., 1993) by a broken power-law
model with slope $\sim 1.85$ for $S \lae 2.2\times 10^{-14}$ \ecs and
$\sim 2.6$ for $S \gae  2.2\times 10^{-14}$ \ecs.  
At brighter fluxes a good agreement is found with the surface density 
previously obtained using data from the {\it Einstein} Observatory 
(Primini et al., 1991; Gioia et al., 1990).

Spectroscopic observations of the  ROSAT sources along with the results
obtained by the {\it Einstein} Observatory Extended Medium  Sensitivity
Survey identification program (EMSS; Gioia et al., 1990; Stocke et al.,
1991; Maccacaro et al., 1994) have allowed us to clarify the nature
and composition of the sources which shine in the soft X-ray sky.

Broad Line AGNs constitute the majority of the soft X-ray sources at
the fluxes actually sampled: they represent $\sim 50 \%$ of the sources
in the EMSS ($<z> \simeq 0.4$) and $\sim 50-60 \%$ of the ROSAT sources with 
$S \gae 5-6\times 10^{-15}$ \ecs ($<z> \simeq 1.5$; Shanks et al., 1991; Boyle
et al., 1993;1994;1995; Jones et al., 1997).

Clusters of galaxies are the next most numerous class of extragalactic
sources in bright X-ray surveys.  At  $S \geq 1\times 10^{-13}$ \ecs , 
for example, they represent $\sim 13 \%$ of the sources in the EMSS. 
However they become less important as one moves to fainter fluxes
(Rosati et al., 1995).

An important minority ($\sim 10\%$) of the ROSAT sources are
spectroscopically identified with  X-ray luminous galaxies, consisting 
of both early-type  and Narrow Emission Line Galaxies (Griffiths et al.,
1995; 1996). According to their cosmological evolution properties
(Griffiths et al., 1996) these objects could become increasingly more
important at fainter fluxes and could constitute a significant fraction
of the largely unidentified sources present at the ROSAT deep surveys
flux limit.  The real nature of these objects (obscured AGN ?
starburst galaxies ?) is at the moment unclear.

Based on studies on the cosmological properties of the known classes of
objects it is now evident that their combined contribution to the soft
CXB is approaching 100 per cent (see e.g. Broad Line AGN: Boyle et al.,
1993;1994; Jones et al., 1997; clusters of galaxies; Rosati et al.,
1995; X-ray luminous galaxies:  Boyle et al., 1995; Griffiths et al.,
1995;1996).

In contrast, our knowledge of the nature and composition of the sources
which produce the 2-10 keV CXB, closer to where the bulk of the energy
density resides, is quite scanty.  Before ASCA (acronym for {\it
Advanced Satellite for Cosmology and Astrophysics}), the surveys in
this energy range were made using passively collimated X-ray
detectors.  These instruments were able to measure accurately the CXB
spectrum (which is remarkably well fitted by a thermal bremsstrahlung
model with a temperature of the order of 40 keV, see Marshall et al.,
1980; Gruber et al., 1992) but, because of the limited spatial
resolution, they were only able to resolve the brightest X-ray sources
which represent only a small fraction ($< 5\%$) of the CXB.

The only statistically complete and large sample of X-ray  sources in
the 2-10 keV  energy band was obtained  from the A-2
experiment on HEAO-1 (Piccinotti et al., 1982).  This sample is
composed of 85 sources (excluding the LMC and SMC), 
brighter than  $\sim  3.1\times 10^{-11}$ \ecs, found in $\sim 8.2 \ sr$ of
the sky at $|b| > 20^o$.  Of the 62 sources of extragalactic origin
about  half are AGN and  half are cluster of galaxies.

Source counts down to a flux of $\sim  8\times 10^{-12}$ \ecs were obtained 
by Kondo et al., 1991 using data from the {\it Ginga} satellite.  
A small sample of 11 sources was derived from the analysis of 383 square degrees.

The {\it Ginga} fluctuations analysis (Warwick and Stewart, 1989;
Butcher et al., 1997) extend the study of the number-flux relationship
to a flux of the order of  $\sim 10^{-13}$ \ecs.  However these studies
are model dependent (the spectral and clustering properties of the
sources have to be assumed)  and do not give direct indications on the
intrinsic nature of the X-ray emitters which are responsible of the CXB.

The so called ``spectral paradox",  i.e. none of the single classes of
known X-ray emitters is characterized by an energy spectral
distribution similar to that of the CXB, further complicates the
situation.

Given this frustrating situation, a  method to derive the contribution
to the hard CXB of the different classes of X-ray sources through
population synthesis models was developed (i.e. by using their average
broad-band spectral properties folded with their cosmological evolution
properties determined in the soft X-ray energy band). Among the
proposed classes of X-ray emitters that could be important for the
production of the hard CXB there are reflection dominated AGN (Fabian,
1989; Zdziarski et al., 1993 and reference therein), strongly absorbed
AGN (Setti and Woltjer, 1989; Madau Ghisellini and Fabian, 1994;
Comastri et al., 1995) and starburst galaxies (Griffiths and Padovani,
1990).

It is clear that a direct measurement, through selection of a sample
of hard X-ray emitters, is therefore crucial to test competing 
models.

The ASCA  satellite is carrying the first imaging instrument in
the 2-10 keV energy band.  This gives us the possibility to determine
the surface density of hard X-ray selected sources to $\lae
10^{-13}$ \ecs  directly from the source counts and to clarify the
nature of the X-ray emitters which are responsible of the hard CXB
through optical spectroscopic follow up observations.

In this paper we report preliminary results on a medium survey program
in the 2-10~keV energy range using data from the GIS2 instrument
onboard the ASCA observatory.  The paper is organized as follows.  
In section 2 we discuss the criteria for the selection and construction of
the images used, for the selection of the sources and for
the definition of the sky coverage.  
In section 3 we report the main result obtained so far:  the number
counts relationship, derived {\bf directly from source counts}, down to
a flux limit of $\sim 6.3\times 10^{-14}$ \ecs.  We compare this
Log(N$>$S)--LogS with previous results from HEAO-1 and {\it Ginga}, with
new results obtained from other groups using ASCA deep survey fields
and with the extrapolation of the Log(N$>$S)--LogS from the soft X-ray
energy band.  Finally in section 4 a discussion and conclusions are
presented.
 
\section{Selection Criteria and Data Analysis}

ASCA  (Tanaka, Inoue and Holt, 1994; Serlemitsos et al., 1995) was
launched on 1993 February 20 by the Japanese Institute for Space
Astronautical Science.  The focal plane instrumentation consists of
two  Solid-State Imaging Spectrometers (hereafter SIS0 and SIS1 or in
general SIS) and two Gas Imaging  Spectrometers (hereafter GIS2 and
GIS3 or in general GIS), simultaneously operating.

For the purpose of this survey we have decided to use the GIS2 
instrument for the following reasons:

a) the GIS has a circular field of view of $\sim 50^{\prime}$ diameter,
$\sim 4$ times larger than the SIS field of view of $\sim 22^{\prime}
\times 22^{\prime}$ (when used in 4 CCD mode).  
Furthermore only a limited number of ASCA observations were made with the 
SIS in the 4-CCD mode option;

b) for energies above $\sim 5$ keV the effective area of the GIS
is larger than the effective area of the SIS;

c) the GIS2 detector has a larger area uncontaminated from a high
Non-X-Ray background than the GIS3 (see the Announcement of Opportunity
EAO\_1; ASCA guest investigator programme).
 
\subsection{ASCA Field Selection and GIS2 Cleaning Parameters}

The ASCA public data considered here were extracted from the 
public archive (as of February 14, 1996).

We have excluded from the survey those fields centered on:

1) $|b| < 20$

2) targets in the Large and Small Magellanic Clouds

3) very bright (see section 2.4) 
and/or extended targets, e.g.  supernovae remnants, very rich and 
nearby clusters of galaxies; 

4) groups and/or associations of targets, e.g. groups of nearby 
galaxies, star clusters or stellar association.

All these selection criteria were applied to avoid regions of high
Galactic absorption and high stellar density, to dismiss images
where the background map (see Section 2.2)
could not be produced reliably and to prevent
including in the survey sources not truly serendipitous (e.g.
physically related to the target).  We have also excluded from the
analysis those images with an exposure time (after the data cleaning,
see below) less than 10000 s since their sensitivity is so poor that
they do not significantly contribute to the sky coverage of the
survey.  Finally in the case of two or more overlapping images we have
retained only the deepest one.

A total of 87 GIS2 fields (listed in Table 1) 
survive to these selection criteria.

Image preparation has been performed using version 1.2 of the XSELECT
software package and version 3.2 of FTOOLS.  Good time intervals were
selected by applying standard cleaning criteria, the most relevant
being a magnetic cut-off rigidity threshold of 6 (COR\_MIN $>$ 6) and a
minimum elevation angle of 6 (ELV\_MIN $>$ 6).  We have also excluded
from the analysis periods of time that are within 60 s from the South
Atlantic Anomaly passage and within 100 s from the day/night
transition passage.  Finally, images were extracted in the 2-10~keV
energy range, in detector coordinates and considering only the events
within 20 arcmin from the detector center. The pixel size of these
images is $15^{\prime\prime} \times 15^{\prime\prime}$.

\subsection{Source Detection and Selection}

To localize and characterize the faint sources contained in the selected 
fields we need to estimate the underlying background as precisely as possible.
A detailed description of the properties of the GIS2 background is 
reported in Kubo et al., 1994 and Ikebe et al., 1995.
In brief the GIS2 background consists of  Non X-ray Background (NXB) and 
Cosmic X-ray Background (CXB); the first one increases from the center to the 
edge of the detector while the latter one shows an opposite behavior.

After a detailed radial and azimuthal analysis of the background
properties in 30 GIS2 images (which do not show evident X-ray
sources) we reached the conclusion that the GIS2 background  has a
structure that is reproduced from image to image.
For these reasons we have decided to use the Blank-Sky event files, 
which include both the NXB and CXB. 
These Blank-Sky event files have been constructed by the GIS team summing  
a total of 15 separate pointings of blank fields
\footnote{Even if these observations do not contain bright X-ray sources, 
we found evidence of 3 faint sources; these sources have been averaged 
out by substituting mean values from surrounding areas.}
and  have been made publicly available.

In order to match as closely as possible the selection criteria used to 
extract the survey images we have extracted Blank-Sky images (hereafter 
BSI) in the 2-10 keV energy band including only the events with a magnetic 
cut-off rigidity threshold above 6.

We have then applied the following multi-step procedure to each 
image (hereafter {\it ima}): 

a) a first estimate of the background statistics in {\it ima} has been
determined in 3 boxes (each of size $5^{\prime} \times 5^{\prime}$) 
which do not
contain evident X-ray sources. Median values of the mean and standard
deviation from these 3 regions have been considered ($\overline{cts}$,
$\sigma$);

b) $\overline{cts}$ and $\sigma$ have been used to define the region 
(hereafter $reg_{back}$) in {\it ima} where the counts (per pixel) are 
$\leq \overline{cts} + 2\sigma$; 

c) this region has been used to produce a normalized version of the 
background map (hereafter {\it back}) by rescaling the BSI according to:

$$ back = BSI \times {Total\ Counts\ in\ ima\ from\ reg_{back}\over Total\ 
Counts\ in\ BSI\ from\ reg_{back}};$$

We have preferred to normalize in this way the exposure map, rather than by 
using the exposure time, because from our analysis of the 30 GIS2 images 
we have found that the background rate can significantly change from image to 
image in its intensity but not in the overall structure across the field. 
 
d) {\it ima} and {\it back} have been then smoothed with a bidimensional 
Gaussian with $\sigma = 1^{\prime}$, which is comparable to the ``core" 
of the (XRT+GIS2) PSF; 

e) the smoothed version of {\it back} has been subtracted from the smoothed 
version of {\it ima} in order to obtain a background-subtracted image 
(hereafter {\it ima-back}); 

f) contour plots in {\it ima-back} have been used to localize the
contiguous pixels (hereafter {\it peaks}) which are 0.5$\sigma$ above
the local mean.
The value of 0.5$\sigma$ has been determined empirically and
ensure completeness above the source detection threshold of 3.5 used (see
below);

g) net counts, in a circle of 2 arcmin radius around each of these peaks, 
are computed by subtracting the background counts accumulated in {\it back} 
from the total counts accumulated in {\it ima};  

h) sources are accepted as real if the signal-to-noise ratio (S/N), 
defined as [(net counts / (net counts + background)$^{1/2}$)], is greater 
than 3.5.    

Excluding the sources related to the targets and the targets themselves,  
a total of 60 X-ray sources satisfy the criteria for the inclusion 
in the survey.

In Table 2 (not reported in the pre-print version) we report the basic data for the 60 sources in the sample.
In particular we list the name, the ASCA ROR where the source was
found, celestial (J2000) coordinates, the signal-to-noise ratio, the
corrected flux (and its error) and the Galactic $N_H$ along the line of
sight.  We note that the absolute accuracy of the ASCA attitude
solution is of the order of 2 arcmin.

\subsection{Correction for vignetting and PSF}

In order to obtain the correct flux of the sources we must now 
consider the position dependent sensitivity of the GIS2 detector.  
In particular we must take into account:
a) the fraction of the source photons falling outside the 
2 arcmin circle radius;
b) the variation of the sensitivity of the detector as a function of 
the distance from the optical axis due to vignetting and 
c) the variation of the (XRT+GIS2) PSF across the field of view.
These effects are taken into consideration by the FTOOLS task ASCAARF 
which is able to produce a position-dependent PSF-corrected effective area 
of the (XRT+GIS2) combination.
In essence the input of ASCAARF is the position $x,y$ (in detector coordinates) 
and the dimension of the source extraction region $dr$ (in our case a circle 
with a radius of 2 arcmin).

Using ASCAARF (version 2.64) we have produced a raster scan of the
effective area values across the GIS2 field of view and used them
(through spectral simulation with XSPEC) to obtain the count rate to
flux conversion factors ($f_{xy}$) relative to each position $x, y$ in
the detector.
In the spectral simulations we have used a power law model with energy
index equal to 0.7, filtered by a Galactic absorbing column density of
$3 \times 10^{20}$ cm$^{-2}$.  The count rate to flux conversion
factors are a very weak function of the Galactic absorbing column
density along the line of sight (which ranges from $10^{20}$ cm$^{-2}$
to $8 \times 10^{20}$  cm$^{-2}$ for the present sample) and are
accurate to $\pm 15\%$ for all the energy spectral indices in the range
0.5--1.0.
The flux of each source has then been computed and is listed in Table 2.

\subsection{Sky Coverage}

The sky coverage (i.e., the area covered as a function of the flux
limit) of the present survey has been determined in the following way.
At each position of a raster scan of a given image we have used the
corresponding background map to compute the total counts that a source
should have had to be detected at the 3.5 S/N level.  These counts have
been then converted to count rate and then flux using the same
conversion factors ($f_{xy}$) used for the sources.

The area associated with the target has been excluded from the sky 
coverage computation; to this purpose we have considered as target 
area the circle defined by the location where the expected counts from 
the pointed source are $\geq 15\%$ of those of the background in the 
corresponding background map.
The circles associated to the targets range between 6 arcmin to 10 arcmin
radius. In the case of circles greater than 12 arcmin radius
we have excluded the whole field from the survey.

An integral representation of the sky coverage is reported in figure 1.

\begin{figure}
\figurenum{1}
\plotone{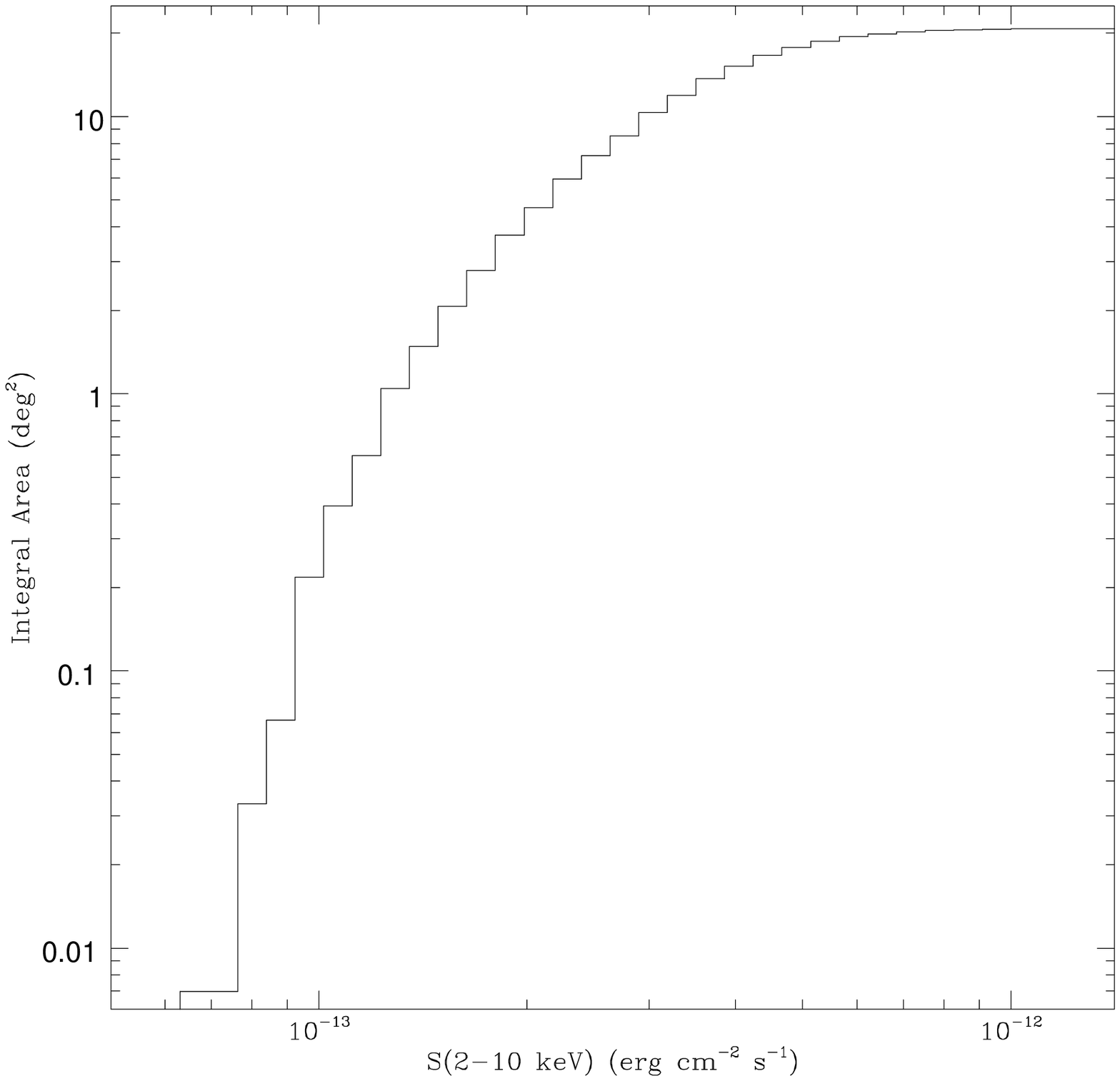}
\caption[Fig.1] {Integral sky-coverage of the present survey 
(see text for details).}
\end{figure}

The raster scan used to compute the sky coverage allows us to estimate
the number of expected spurious sources in the survey.  The number of
background counts in {\it back} inside a circle of 2 arcmin radius is a
function of a) the image exposure time and b) the circle position in
the image.  The background counts range between 10 and 15 counts for an
image with an exposure time of $\sim 10$ ksec, between 35 and 50 counts
for an image with an exposure time of $\sim 40$ ksec and between 65 and
85 counts for an image with an exposure time of $\sim 70$ ksec.
Considering the extremes of these values, the requirement of S/N = 3.5
implies that a minimum of 19 (39) net counts have to be recorded on top
of the 10 (85) background counts.  For a Poisson distribution the
probability of observing 29 (124) total counts or more when 10 (85) are
expected is $7.6\times 10^{-7}$ ($4.3\times 10^{-5}$).  We note that
the 4 $\sigma$ level in Gaussian statistics is $\sim 6\times 10^{-5}$.
The number of resolution elements of this survey is $\sim 6000$ (total
area divided by the detection cell area); thus the number of spurious
sources expected  is $<0.3$.

\section{Results}

In figure~2 we show (open circles) a non-parametric representation of
the number-flux relationship, obtained by folding the  sky
coverage with the flux of each source.
Because we are primarily interested in the extragalactic number-flux relationship we have excluded from the computation  two sources 
(a0341-4353 and a1410+5215) which are suspected to be stars.
In the same figure we also report a parametric representation (solid
line) of the LogN($>$S)--LogS, obtained by applying the maximum
likelihood method to the unbinned data (see Gioia et al.,1990 for
details on these two representations).
The fit has been performed from a flux of $\sim 6.3\times 10^{-14}$
\ecs (the faintest detectable flux) to a flux of $\sim 10^{-11}$ \ecs.
For fluxes brighter than this limit we may not be complete since most
of the ``bright" X-ray sources were chosen as target of the
observations and then excluded, by definition, from the survey.
However we note that the space density of sources with flux greater
than $\sim 10^{-11}$ \ecs is such that less than 0.2 sources are
expected in this survey.  We also note that less than 1 source is
expected with  flux between $\sim 6.3\times 10^{-14}$ \ecs and $\sim
1.1\times 10^{-13}$ \ecs (the flux of the faintest detected source).

\begin{figure}
\figurenum{2}
\plotone{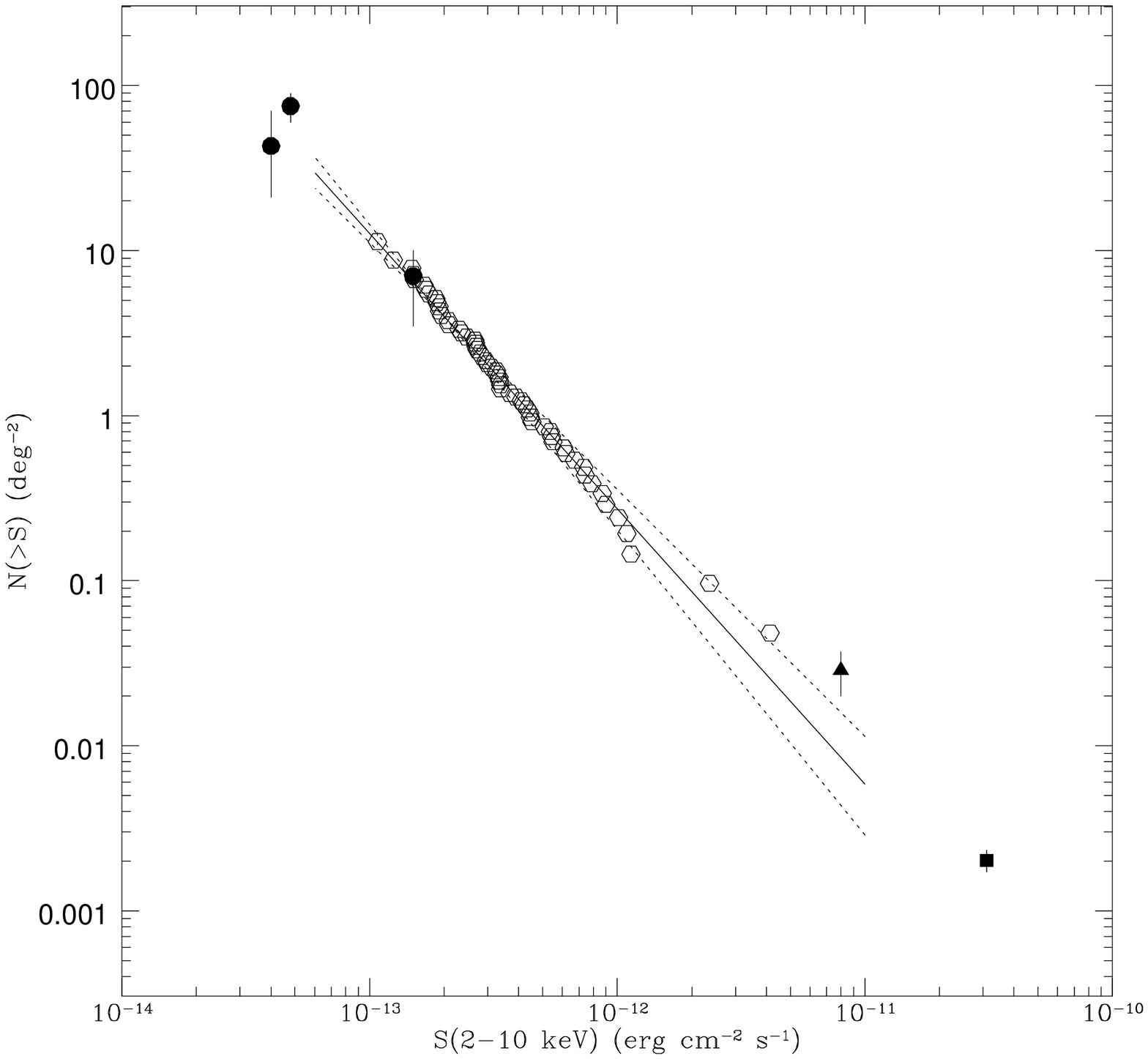}
\caption[Fig.2]{ Non-parametric (open circles) representation of the
number-flux relationship obtained with the selected sample. The
parametric representation (solid line) is well described by a power law
model, N($>$S) = K $\times S^{-\alpha}$, with best fit value for the
slope of $\alpha = 1.67\pm 0.18$ and  K = $2.85 \times 10^{-21}$
deg$^{-2}$.  The dotted lines represent the $\pm 68\%$ confidence
intervals on the slope.  The filled square represents the extragalactic
surface density in the Piccinotti et al., 1982 sample, corrected for
the 20\% excess of sources due to the local superclusters (as estimated
by Comastri et al., 1995).  The filled triangle represents the results
obtained by Kondo et al., 1991 using Ginga data.  The filled dots are
preliminary results in ongoing ASCA survey program conducted on a more
limited sky area (Inoue et al., 1996; Georgantopoulos et al., 1997)}
\end{figure}

The LogN($>$S)--LogS can be described by a power law model 
N($>$S) = K $\times S^{-\alpha}$ with best fit value
for the slope of $\alpha = 1.67$; the $68\%$ and $90\%$ confidence
intervals are equal to [1.49;1.84] and [1.38;1.96] respectively 
(see Table 3). 
The normalization K is determined by rescaling the model to the actual
number of objects in the sample and, in the case of the ``best" fit
model, is equal to K = $2.85 \times 10^{-21}$ deg$^{-2}$.  
The two dotted lines in figure 2 represent the $\pm 68\%$ confidence intervals
on the slope.  A consistent ``best" fit model is obtained if we 
consider only the 26 sources detected with a S/N ratio greater than 5 and 
the corresponding sky coverage 
( $\alpha = 1.71\pm 0.26$; K = $8.28 \times 10^{-22}$ deg$^{-2}$).

The filled square at $\sim 3\times 10^{-11}$ \ecs re\-pre\-sents the
surface density of the extragalactic population in the  Piccinotti et al., 
1982 sample, corrected for the 20\% excess of sources due 
to  the local superclusters (as estimated by Comastri et al., 1995).

The filled triangle at $\sim 8\times 10^{-12}$ \ecs represents the
surface density of X-ray sources as determined by Kondo et al., 1991 
using a small sample of 11 sources extracted from the Ginga high Galactic
latitude survey.

Finally, the surface densities represented by the filled dots at 
$\sim 2\times 10^{-13}$ \ecs and $\sim 4\times 10^{-14}$ \ecs are preliminary
results from an ongoing ASCA survey programs conducted on a more limited
area of sky (as reported in Inoue et al., 1996), while the filled dot 
at  $\sim 5\times 10^{-14}$ \ecs is a recent results obtained by 
Georgantopoulus et al., 1997 using 3 deep ASCA GIS observations. 

Consistent results (at a flux limit of $\sim 6\times 10^{-14}$ \ecs) 
are obtained using data from the BeppoSAX deep surveys (P. Giommi, 
private communication). 

Figure~2 shows that we have extended the description of the (2-10
keV) number-flux relationship by  a factor $\sim 450$  with respect to
the Piccinotti et al., (1982) determination, with a sample of similar size.

We can now evaluate, directly from the source counts, the 
contribution of the resolved sources to the 2-10 keV CXB.  
This contribution is the ratio between the background surface 
brightness and the emissivity of the objects under study, which is 
obtained via integration of the observed LogN($>$S)--LogS.

In the 2-10~keV energy range the spectrum of the CXB can be described
by a power law representation with energy spectral index  equal to 0.4
and normalization equal to 8 keV cm$^{-2}$ s$^{-1}$ sr$^{-1}$
keV$^{-1}$ (see the review of Hasinger, 1996), implying a background
surface brightness of $\sim 32.87$ keV cm$^{-2}$ s$^{-1}$ sr$^{-1}$.

The emissivity of the resolved objects, $I_{obj}$, is given by
 $$ I_{obj} = \int_{S_{min}}^{\infty} S \times N(S) dS$$ 
where S
is the source flux, N(S)dS is the differential representation of the
number-flux relationship and $S_{min}$ is the limiting flux considered.

Using the LogN($>$S)--LogS best fit model (and the $\pm 68$ \% confidence 
interval) and 
$S_{min} = 6.3\times 10^{-14}$ \ecs 
(the flux limit of the survey) we obtain $
I_{obj} = 8.8^{+0.6}_{-0.2} $ keV cm$^{-2}$ s$^{-1}$ sr$^{-1}$, which
represents about 27\% of the 2-10 keV CXB.

We note that our determination of the total intensity produced by the
detected sources is {\it independent} of the intensity of the CXB and
therefore  $I_{obj}$ is not affected by the possible uncertainties
which could affect the estimate of the CXB.

The LogN($>$S)--LogS best fit model implies the saturation of the 2-10 keV
CXB at a flux of $\sim 8.7\times 10^{-15}$ \ecs.  Therefore deeper
X-ray surveys in this energy range (e.g. the SAX, Jet-X and XMM deep
surveys) should find a flattening of the number-flux relationship in
the range  $\sim 8.7\times 10^{-15}$ -- $\sim 6.3\times 10^{-14}$
\ecs, i.e. within a factor $\sim 10$ from the present flux limit.

\section{Discussion and conclusion}

It is clear that a decisive and final understanding of the real 
nature of the hard X-ray sources in this sample must await their 
spectroscopic identification.

In the meantime we can infer useful information on their nature by
comparing the measured 2-10 keV LogN($>$S)--LogS with that extrapolated
from the soft energy band and with the predictions obtained from
population synthesis models.

\begin{figure}
\figurenum{3}
\plotone{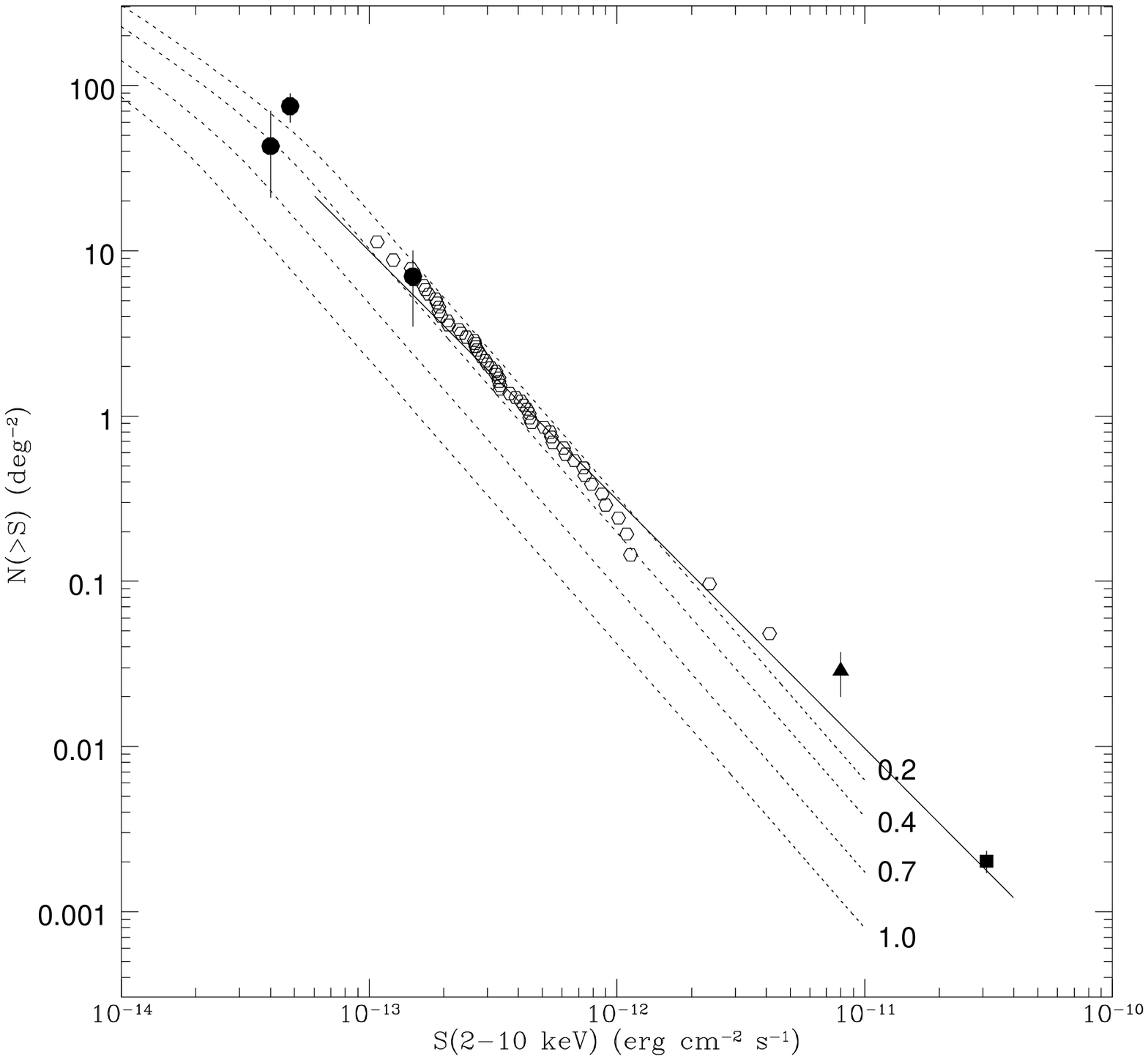}
\caption[Fig.3]{Comparison of the derived LogN($>$S)-LogS (open circles)
with the soft (0.5 -- 2.0 keV) LogN($>$S)-LogS (Hasinger et al., 1993)
extrapolated in the 2--10 keV energy band assuming a power law spectral
model with energy indices equal to 1, 0.7, 0.4 and 0.2 (dotted lines) The
solid line represents the prediction of the Comastri et al. (1995)
model.}
\end{figure}

In figure 3 we display the LogN($>$S)--LogS obtained in this work along
with:  a) the soft (0.5-2.0 keV) LogN($>$S)--LogS (Hasinger et al.,
1993) extrapolated to the 2-10 keV energy band assuming a power law
spectral model with energy indices equal to 1, 0.7, 0.4 and 0.2 (dotted
lines); and b) the prediction of the Comastri et al., (1995) model which
is based on the Unification scheme of the AGNs (solid line).

We will discuss these two comparisons and their implications in 
turn.

The first comparison clearly shows a well known problem (see e.g.
Warwick and Stewart, 1989): 
if we  estimate the hard X-ray LogN($>$S)--LogS by
extrapolating the soft X-ray LogN($>$S)--LogS and using the mean
spectral properties of the sources, as determined in the soft
(0.5 -- 2.0 keV) energy band ($\alpha \sim 0.7 - 1.2$, 
see e.g. Almaini et al., 1996),  we fail to reproduce it.  
In particular we predict a surface density, at a flux of the order 
of $10^{-13}$ \ecs, which is at least a factor $\sim 2$ less than the 
observed one.

If the same sources are responsible for both the soft and the 
hard  X-ray LogN($>$S)--LogS, their average spectral energy index should
be of the order of 0.3. 
 
Since at fluxes brighter than $\sim 10^{-13}$ \ecs the principal
contribution to the source counts in the soft energy band is due to
broad line AGNs (Seyfert 1 and QSOs) we should see a drastic change of
their mean spectral properties going from the soft to the hard energy
band.  This hypothesis seems to be contradicted by the observations
(see e.g.  Comastri et al., 1992; Williams et al., 1992; Matsuoka and
Cappi, 1996; Cappi et al., 1997).

An alternative hypothesis is that, in the hard energy band,  we 
are selecting a population of sources which is actually undersampled 
in the soft X-ray surveys (e.g. the heavily absorbed AGNs in the Madau, 
Ghisellini and Fabian, 1994 and Comastri et al., 1995 models).
In this respect it is worth noting the discovery (Ohta et al., 1996) of 
a type 2 QSOs at z=0.9 in an ASCA SIS observation of the Lynx field.

In particular the model of Comastri et al., (1995) is based on the 
X-rays properties of the Unification Scheme of AGNs:  
according to this model a population of absorbed and
unabsorbed AGNs, when folded with the cosmological evolution properties
determined in the soft X-ray energy band, is able to reproduce the
shape and intensity of the CXB  from several keV to $\sim 100$ keV.  
As shown in figure 3 the predicted 2-10 keV LogN($>$S)-LogS of 
Comastri et al. (1995) (reported as a solid 
line) in the flux range $\sim 6\times 10^{-14}$ - $\sim 4\times 10^{-11}$ 
\ecs is in very good agreement with our determination.

If this model is correct we can predict that $\sim 52$ AGNs and $\sim 8$
Clusters of Galaxies should be present among the 60 sources of the
present sample.  The expected composition of the AGNs population 
in this survey as a function
of the intrinsic $N_H$ is reported in Table 4.
The population of AGNs X-ray emitters with intrinsic $N_H$ in the range  
$10^{23} - 10^{24}$ cm$^{-2}$ 
is the most interesting one:  according to 
Comastri et al., 1995 they should produce the largest fraction of the CXB 
in the 2 -- 100 keV energy range and should be practically invisible 
in  soft X-ray surveys even at the faintest fluxes reachable with 
current and future X-ray missions.

We are planning to extend  the survey by a factor of $\sim 2-3$ 
by analyzing all the available and suitable public GIS2 data.
In the meantime we have started  a program to identify the 
optical counterparts of the X-ray sources. 
The main problem with the spectroscopic follow up is the 
absolute accuracy of the ASCA attitude solution ($\sim 2^{\prime}$). 
Fortunately ROSAT PSPC data are available for most of the 
selected GIS2 fields, allowing us to greatly reduce the X-ray errors circle 
for many of the X-ray sources associated with unabsorbed AGN and with 
clusters of galaxies.
A comparison of the ASCA  and ROSAT  spectra
of the selected sources  with that of the CXB is in progress and will 
be published elsewhere.

\acknowledgments

We are grateful to A.Comastri, M.Cappi, G.Ghisellini and A. Wolter for stimulating discussions.
We thank the anonymous referee for the carefull reading of the manuscript 
and for useful comments.

\onecolumn

\begin{deluxetable}{lcccc}
\tablenum{1}
\tablecolumns{5}
\tablewidth{0pc}
\tablecaption{List of ASCA fields used}
\tablehead{
\colhead{Target Name}    & \colhead{ROR}    &
\colhead{$\alpha$ \tablenotemark{a}}    
& \colhead{$\delta$ \tablenotemark{a}}    &  \colhead{$T_{\rm exp}$ \small (s)} 
}
\startdata
 BETA CET    &  21019000 &  00:43:37.0   &    $-$18:03:56 &  19977\nl
 GSGP4       &  92005000 &  00:57:29.8   &    $-$27:37:20 &  36934\nl
 GRB191178   &  22034000 &  01:18:40.2   &    $-$28:39:14 &  27160\nl 
 NGC 507     &  61007000 &  01:23:03.3   &    $+$33:20:00 &  37591\nl 
 NGC 720     &  60004000 &  01:53:19.0   &    $-$13:42:27 &  34213\nl  
 MRK 1040    &  72016000 &  02:28:23.0   &    $+$31:22:15 &  19853\nl 
 PHL 1377    &  72031000 &  02:35:17.9   &    $-$03:57:21 &  31683\nl 
 AO 0235+164 &  71015020 &  02:38:30.0   &    $+$16:32:02 &  24858\nl 
 ABELL 370   &  80010000 &  02:39:58.5   &    $-$01:31:57 &  34812\nl 
 NGC1097     &  61001000 &  02:46:04.6   &    $-$30:18:01 &  36872\nl 
 NGC1313     &  60028000 &  03:18:25.9   &    $-$66:28:24 &  23282\nl 
 NGC1316     &  61002000 &  03:22:25.3   &    $-$37:14:18 &  34007\nl 
 1H0323+022  &  71034000 &  03:26:06.2   &    $+$02:20:43 &  34584\nl 
 EPSILON ERI &  22009000 &  03:33:04.8   &    $-$09:24:12 &  19459\nl 
 QSF3        &  90011000 &  03:41:44.7   &    $-$44:07:06 &  22967\nl 
 MS0353.6-364 &  82042000 & 03:55:47.7   &    $-$36:32:12 &  17864\nl 
 VW HYI      &  31001000 &  04:09:59.1   &    $-$71:21:03 &  12132\nl 
 NGC1614     &  61011000 &  04:33:49.7   &    $-$08:39:08 &  32957\nl 
 PKS0438-436 &  70010000 &  04:40:22.0   &    $-$43:32:29 &  32569\nl 
 NGC 1667    &  71032000 &  04:48:27.0   &    $-$06:23:43 &  38173\nl 
 E0449-184   &  71033000 &  04:51:27.0   &    $-$18:23:20 &  36489\nl 
 MSS0451.6-03 &  81025000 & 04:54:01.2   &    $-$03:05:33 &  52439\nl 
 MS0451.5+02 &  82041000 &  04:54:26.4   &    $+$02:54:58 &  18684\nl 
 NGC 1808    &  71031000 &  05:07:27.7   &    $-$37:35:17 &  32750\nl 
 MKN 3       &  70002000 &  06:15:25.3   &    $+$71:01:53 &  27640\nl 
 0716+714    &  71006020 &  07:20:59.8   &    $+$71:17:46 &  19020\nl 
 YY GEM      &  20002000 &  07:34:57.8   &    $+$31:57:55 &  30441\nl 
 PI1 UMA     &  21018000 &  08:39:41.5   &    $+$65:05:16 &  24336\nl 
 Lynx Field  &  90009010 &  08:49:11.3   &    $+$44:50:08 &  39943\nl 
 IRAS 09104  &  71002000 &  09:13:52.6   &    $+$40:56:14 &  38492\nl 
 GL 355      &  21020000 &  09:32:46.3   &    $-$11:09:37 &  19166\nl 
 NGC 2992    &  71049000 &  09:45:26.2   &    $-$14:22:52 &  25914\nl 
 NGC3079     &  60000000 & 10:01:12.4   &    $+$55:38:07 &  38704\nl 
 3C234       &  71043000 & 10:01:27.8   &    $+$28:45:12 &  16194\nl 
 A963        &  80000000 & 10:16:32.6   &    $+$39:00:12 &  23077\nl 
 NGC 3147    &  60040000 & 10:17:06.4   &    $+$73:23:25 &  36120\nl 
 AD LEO      &  20006000 & 10:19:18.7   &    $+$19:53:44 &  26183\nl  
 IRASF10214+4 &  61003000 & 10:25:03.1   &    $+$47:10:34 &  34854\nl 
 RE1034+39   &  72020000 & 10:35:04.3   &    $+$39:40:29 &  28345\nl 
 NGC\_3310   &  61013000 & 10:38:11.3   &    $+$53:29:19 &  17648\nl 
 LOCKMAN HOLE &  90010020 & 10:52:08.6   &    $+$57:22:39 &  21593\nl 
 A1204       &  82002000 & 11:12:50.6   &    $+$17:32:40 &  27692\nl 
 NGC 3628    &  61015000 & 11:20:31.0   &    $+$13:34:19 &  21616\nl 
 HCG51       &  82028000 & 11:22:06.2   &    $+$24:15:02 &  11694\nl 
 MKW4\_NE    &  82013000 & 12:05:07.2   &    $+$01:57:31  &  19574\nl
 NGC 4203    &  61008000 & 12:15:21.1   &    $+$33:10:22 &  36451\nl  
 NGC\_4418   &  62003000 & 12:26:35.0   &    $+$00:54:59 &  25875\nl 
 MS1224.7+200 &  82043000 & 12:27:04.1   &    $+$19:53:02 &  11533\nl 
 NGC4449     &  62011000 & 12:27:43.0   &    $+$44:05:25 &  45021\nl 
 NGC 4472 NW8 &  60030000 & 12:29:23.1   &    $+$07:54:16 &  20128\nl 
 NGC4643     &  62001000 & 12:42:52.3   &    $+$01:55:56 &  35222\nl 
 NGC4649     &  61005000 & 12:43:54.0   &    $+$11:31:37 &  25447\nl 
 MS1248.7+570 &  72032000 & 12:50:21.8   &    $+$56:49:13 &  31492\nl 
 GP COM      &  32002000 & 13:05:22.3   &    $+$17:58:54 &  29922\nl 
 NGC 4968    &  71039000 & 13:07:28.3   &    $-$23:39:01 &  25755\nl 
 LSS\_LineC-P & 92002070 & 13:13:02.9   &    $+$31:21:36 &  16928\nl 
 A1704       &  81007000 & 13:13:50.6   &    $+$64:38:19 &  18570\nl 
 ABELL 1722  &  81013000 & 13:19:53.5   &    $+$70:05:20 &  39580\nl
 NGC5236     &  61016000 & 13:37:04.3   &    $-$29:52:08 &  10474\nl 
 NGC5252     &  71021000 & 13:38:33.6   &    $+$04:35:08 &  67795\nl 
 CL1358+6245 &  81032000 & 13:59:23.3   &    $+$62:35:28 &  35062\nl 
 K416        &  62007000 & 14:05:51.4   &    $+$50:44:56 &  33977\nl 
 PG 1404+226 &  72021000 & 14:06:07.0   &    $+$22:22:52 &  25639\nl 
 PKS 1404-267 &  81019000 & 14:07:53.0   &    $-$26:59:12 &  34280\nl 
 PG1407+265  &  70024000 & 14:08:54.4   &    $+$26:16:19 &  23344\nl 
 3C295       &  71003000 & 14:11:00.0   &    $+$52:14:40 &  44926\nl 
 MS1426.4+01 &  82044000 & 14:28:46.8   &    $+$01:46:41 &  18734\nl 
 NGC 5695    &  71009000 & 14:37:39.8   &    $+$36:34:09 &  24626\nl 
 HD 129333   &  22012000 & 14:38:16.6   &    $+$64:19:39 &  19008\nl  
 NGC 5846    &  61012000 & 15:06:37.4   &    $+$01:36:39 &  39548\nl 
 3C313       &  71004000 & 15:11:17.3   &    $+$07:54:21 &  41082\nl 
 ARP 220     &  60035000 & 15:34:29.7   &    $+$23:27:42 &  33031\nl 
 PG\_1634+706 & 71036000 & 16:34:25.0   &    $+$70:35:27 &  38496\nl 
 ABELL 2218  &  80001000 & 16:35:56.6   &    $+$66:15:38 &  23913\nl 
 A2219       &  82037000 & 16:39:42.7   &    $+$46:40:27 &  33941\nl 
 NGC6240     &  71022000 & 16:53:13.2   &    $+$02:28:15 &  31539\nl 
 DRACO 3     &  90024000 & 17:10:24.2   &    $+$71:06:07 &  33590\nl 
 PG\_1718+481 & 71037000 & 17:19:38.6   &    $+$48:08:07 &  33194\nl 
 NEP FIELD 1 &  90020000 & 17:59:59.8   &    $+$66:33:48 &  21718\nl  
 PAVO\_GROUP &  81020000 & 20:18:33.1   &    $-$70:51:12 &  30942\nl 
 PKS 2126-158 &  70014000 & 21:29:20.4   &    $-$15:34:58 &  11044\nl 
 MS2137.3-235 &  81022000 & 21:40:29.0   &    $-$23:38:35 &  16794\nl  
 A2440       &  81033000 & 22:23:43.2   &    $-$01:37:15 &  36602\nl 
 PHL 5200    &  72033000 & 22:28:37.7   &    $-$05:13:42 &  13130\nl 
 IC1459      &  60005000 & 22:57:15.6   &    $-$36:27:23 &  27002\nl 
 DI\_PEG     &  22002000 & 23:32:21.8   &    $+$15:03:11 &  33458\nl 
 A2670       &  82049000 & 23:54:30.2   &    $-$10:19:42 &  27628\\ 
\enddata
\tablenotetext{a}{Equatorial coordinates; equinox J2000}
\end{deluxetable}

\begin{deluxetable}{lcccc}
\tablenum{3}
\tablecolumns{3}
\tablewidth{0pc}
\tablecaption{The Log(N$>$S)-Log(S) Fit Parameters}
\tablehead{
\colhead{}     &\colhead{$\alpha$}  &\colhead{K {\small $(deg^{-2})$}}
}
\startdata
$-$90\%         &1.38  &$1.04\times 10^{-17}$\nl
$-$68\%		&1.49 	&$4.45\times 10^{-19}$\nl 
best fit 	&1.67 	&$2.85\times 10^{-21}$\nl
$+$68\%		&1.84 	&$1.52\times 10^{-23}$\nl
$+$90\%         &1.96  &$4.76\times 10^{-25}$\\
\enddata
\end{deluxetable}

\begin{deluxetable}{lcccc}
\tablenum{4}
\tablecolumns{5}
\tablewidth{0pc}
\tablecaption{AGNs predicted by Comastri et al., 1995 model}
\tablehead{
\colhead{$Log \ N_{H}$}    & \colhead{ predicted AGN}\\   
\colhead{\small (cm$^{-2}$)}  & \colhead{} 
}

\startdata
$<21$ & 26\nl
$21-22 $ & 7\nl
$22-23 $ & 11\nl
$23-24 $ & 8\nl
$24-25 $ & $<1$\\
\enddata
\end{deluxetable}

\end{document}